# Bandgap Engineering for Efficient Perovskite Solar Cells Under Multiple Color Temperature Indoor Lighting

Miqad S. Albishi [1+], Faisal I. Alabdulkarem [2+], George Perrakis [3+], Tariq F. Alhuwaymel [1+], Ala H. Sabeeh [4-5] , Abdullah S. Alharbi [1], Naif R. Alshamrani [1], Ibrahim H. Khawaji [4-5], Nikolaos Tzoganakis [6], Majed M. Aljomah [1] , Dimitris Tsikritzis [6], Sami A. Alhusaini [1], Abdullah Aljalalah [1] , Kadi S. AlShebl [1], Ali Alanzi [1] , Abrar Bin Ajaj [7], Fay M. Alotaibi [1], Hamad Albrithen [7-10] Konstantinos Petridis [8], Maria Kafesaki [3,9], Emmanuel Kymakis [6], George Kakavelakis [8,*], Essa A. Alharbi [1,*].

[1] Microelectronics and Semiconductor Institute, King Abdulaziz City for Science and Technology (KACST),Riyadh, 11442, Saudi Arabia

[2] Sustainable Energy Technologies Center, College of Engineering, King Saud University, Riyadh, Saudi Arabia

[3] Institute of Electronic Structure and Laser (IESL), Foundation for Research and Technology – Hellas (FORTH), 70013 Heraklion, Crete, Greece

[4] Energy, Industry, and Advanced Technologies Research Center, Taibah University, Madinah, Saudi Arabia

[5] Department of Electrical Engineering, College of Engineering, Taibah University, Madinah, Saudi Arabia.

[6] Department of Electrical & Computer Engineering, Hellenic Mediterranean University (HMU), Heraklion 71410, Crete, Greece.

[7] Physics and Astronomy Department, College of Science, King Saud University, Riyadh 11451, Saudi Arabia.

[8] Department of Electronic Engineering, School of Engineering, Hellenic Mediterranean University, Romanou 3, Chalepa, Chania, Crete GR-73100, Greece

[9] Department of Materials Science and Engineering, University of Crete, 70013 Heraklion, Crete, Greece

[10] King Abdullah Institute for Nanotechnology, King Saud University, Riyadh 11451, Saudi Arabia.

[+]These authors contributed equally to this work.

*Corresponding authors: ealharbi@kacst.gov.sa and kakavelakis@hmu.gr

**Abstract**

Perovskite indoor photovoltaics (PIPVs) are emerging as a transformative technology for low-light intensity energy harvesting, owing to their high-power conversion efficiencies (PCEs), low-cost fabrication, solution-processability, and compositionally tunable band gaps. In this work, methylammonium-free $Cs_xFA_{1-x}Pb(I_{1-y}Br_y)_3$ perovskite absorbers were compositionally engineered to achieve band gaps of 1.55, 1.72, and 1.88 eV, enabling matching the spectral photoresponse with the indoor lighting. Devices based on a scalable mesoscopic n-i-p architecture were systematically evaluated under white LED illumination across correlated color temperatures (3000-5500 K) and light intensities from 250 to 1000 lux with active area of 1 $cm^2$. The 1.72 eV composition exhibited the most promising performance across different light intensities and colors, achieving PCEs of 35.04% at 1000 lux and 36.6% at 250 lux, with a stable device operation of over 2000 hours. On the other hand, the 1.88 eV band-gap variant reached a peak PCE of 37.4% under 250 lux (5500 K), however performance trade-offs were observed across the different color lights LEDs. Our combined experimental and theoretical optical-electrical simulations suggest that decreasing trap-assisted recombination in wide-bandgap compositions may further improve PIPV performance across the different illumination conditions. In contrast, devices with 1.55 eV band gap underperformed in such conditions due to suboptimal spectral overlap and utilization. These findings establish bandgap optimization and device architecture as key design principles for high-efficiency, stable PIPVs, advancing their integration into self-powered electronic systems and innovative indoor environments.

**Introduction**

Perovskite solar cells (PSCs) are emerging as a transformative technology within the field of photovoltaics, attracting substantial interest due to their impressive advancement in power conversion efficiency (PCE), which has reached 27% under standard outdoor solar illumination.[1–2] Beyond outdoor applications, PSCs are increasingly recognized as a viable candidate for indoor photovoltaics (IPVs), owing to their tunable energy bandgap ($E_g$), high specific power output, and exceptionally high absorption coefficients.[4,6] This adaptability enables effective energy harvesting under various indoor lighting conditions, including light-emitting diode (LED) lamps and compact fluorescent lamps (CFLs).[3-6] Nevertheless, it is essential to note that the Shockley-Queisser (SQ) limit applicable under indoor lighting conditions diverges from the outdoor limit of approximately 33% for a bandgap of 1.34 eV. For IPVs, the SQ limit exceeds 50%, requiring $E_g$ in the range of 1.7-2.0 eV to align with the narrowband emission spectrum (400-700 nm), a characteristic of indoor lighting.[9-10,26] Perovskite indoor photovoltaics (PIPVs) due to their high PCE in such illumination conditions offer opportunities for a wide range of applications, including powering internet of things (IoT) devices, wireless sensors, wearable health monitors, and smart home actuator.[7–8] One promising strategy for achieving an optimal $E_g$ of 1.7-2.0 eV is compositional engineering, specifically through the manipulation of the X-site in the ABX$_3$ perovskite structure.[13] Numerous studies have focused on varying bromide content (i.e., the iodide-to-bromide ratio in the perovskite composition), resulting in a PCE of 27% with an $E_g$ of 1.61 eV and a PCE of 33% with an $E_g$ of 1.77 eV under indoor LED illumination at 1000 lux.[14-15] Another practical approach involves interface engineering, aimed at reducing defects through surface and bulk passivation methods, leading to reported PCEs exceeding 40%. Specifically, a PCE of 44.72% with a $E_g$ of 1.71 eV represents the highest value reported to date in perovskite IPVs.[20-23] However, such exceptional

PCE results have been achieved using small-area devices (0.09-0.1 cm$^2$),[17] thus emphasizing the necessity of scaling to larger active areas ($\geq$1 cm$^2$) for practical IPV applications, which would provide sufficient power input to drive external devices effectively. Furthermore, most of the previously reported research involves perovskites that incorporate highly volatile A-cations (including the aforementioned), such as methylammonium, while facilitating a wider material $E_g$, compromise device stability. [44-46] Additionally, it is noteworthy that the substantial majority of existing literature on PIPVs is centered on the p-i-n device configuration, which has demonstrated high and stable certified PCEs under both outdoor and indoor illumination.,[46-68] However, there is a lack of robust evidence suggesting that this configuration (p-i-n) presents a scalable solution for the industrialization of PIPVs, particularly when contrasting with the proven scalability of n-i-p device configuration which involves the use of scalable, cost-effective and eco-friendly fluorine-doped tin oxide (FTO) transparent electrode and titanium dioxide (TiO$_2$) electron transporting layer. In contrast the p-i-n devices employ indium tin oxide (ITO) alongside the various limitations associated with, including cost, thermal stability, and indium toxicity.[49-50] Moreover, the extensive characterization and testing of perovskite IPVs in prior studies have predominantly employed warm white LED (WLED) light (3000 K).[30-31] At the same time, indoor environments often encompass a diverse array of lighting conditions.[30-31] Therefore, it is crucial to assess the performance of PIPVs under various color temperatures to ensure consistent operational efficacy for indoor settings. White LEDs are widely utilized as the primary indoor lighting source, principally due to their efficiency and longevity.[18] Their color temperature (CT) ranges from warm (2700-3500 K) to neutral (3500-4500 K) and cool (above 4500 K). Warm light tends to exhibit a redshift, in contrast to the blue shift associated with cool light. These spectral variations inherently influence the performance of any photovoltaic device, including perovskite IPVs.[20]

Herein, this study investigates the influence of varying $E_g$ (1.55 eV, 1.72 eV, and 1.88 eV) of metal halide perovskite materials, specifically through the modulation of the Iodide-to-Bromide (I/Br) ratio within the perovskite composition, on photovoltaic performance under indoor lighting conditions. To enhance stability and avoid the use of volatile A-cations, a methylammonium (MA)-free perovskite formulation incorporating cesium (Cs) and formamidinium (FA) as A-cations was optimized.[70] This approach has not been previously employed for PIPV applications, although it has demonstrated promising results for stable PSCs under outdoor solar illumination. The investigation further encompassed the use of WLED illumination at correlated CTs of 3000 K, 4000 K, and 5500 K, across varying light intensities, to explore the effects of incident light spectrum and perovskite $E_g$. A theoretical framework was established through comprehensive optical full-wave electromagnetic simulations, electrical drift-diffusion modeling, and detailed analysis based on the Shockley-Queisser limit across a spectrum of $E_g$, CTs, and light intensities. This multifaceted approach provided essential insights into the physical mechanisms that constrain the indoor PCE of PIPV under diverse lighting conditions. The combined experimental and simulation results underscore the need to optimize PSC design for consistently high PCEs across various indoor environments. By adjusting the perovskite $E_g$ to 1.88 eV through precise I/Br ratio manipulation, a stable indoor PCE of 37.4% was obtained under illumination of 250 lux and a CT of 5500 K utilizing a scalable and stable n-i-p device configuration, which remains largely unexplored in the context of IPVs. Additionally, a series of characterization techniques was employed to elucidate the critical morphological, compositional, electrical, and optical properties of the optimized materials and devices.

**Results and discussion**

To establish a robust baseline for perovskite composition and device configuration with a focus on long-term stability and upscaling potential, the present study concentrates on the long-term stable formamidinium-cesium-based perovskite composition and the n-i-p mesoscopic PSCs (i.e. FTO/c-TiO$_2$/m-TiO$_2$/Cs$_x$FA$_{1-x}$Pb(I$_{1-y}$Br$_y$)$_3$/SpiroOMeTAD/Au, where $x$=0.10-0.15 and $y$=0.02-0.85). The emission spectra of white light-emitting diodes (LEDs) featuring various CTs and intensities, as well as the spectral power irradiance utilized in this investigation, are depicted in Figures S1 and S2.

Photoluminescence (PL) spectra were systematically collected and analyzed to determine the resulting $E_g$ of those mixed halide perovskites by varying the I/Br ratio. As illustrated in figure 1a, the film, FA$_{0.90}$Cs$_{0.10}$Pb(I$_{0.98}$Br$_{0.02}$)$_3$ exhibited a peak emission wavelength around 800 nm, corresponding to an $E_g$ of 1.55 eV. As the bromine concentration increased, the films FA$_{0.85}$Cs$_{0.15}$Pb(I$_{0.55}$Br$_{0.45}$)$_3$ and FA$_{0.85}$Cs$_{0.15}$Pb(I$_{0.15}$Br$_{0.85}$)$_3$ displayed emission peaks at 720 nm and 662 nm, yielding bandgaps of 1.72 eV and 1.88 eV, respectively. Thus, these results confirm the successful identification of three distinct $E_g$ (1.55 eV, 1.72 eV, and 1.88 eV). [36,63-64]

Subsequent X-ray diffraction (XRD) measurements were performed to elucidate the impact of the I/Br ratio on crystallinity and phase purity across different films. Figure 1b shows a minor peak at 11.6° for samples with 2% and 45% Br, which is associated with the δ- phase perovskite of FAPbI$_3$. [47-48,69-70] Furthermore, another weak peak at 12.6° in the 2% Br and 45% Br films correspond to the PbI$_2$ peak. In contrast, the δ-phase and PbI$_2$ peak are absent in the 85% Br film, indicating a fully reacted final composition with no double phase coexistence. [38,41,51] The very small intensity of the the δ-phase and PbI$_2$ peaks in the 2% and 45% Br-content perovskite films and also their absence in the 85% Br-content perovskite film, indicate a successful incorporation of Br in the perovskite lattice which did not negatively affect the purity and stability of the crystal.

A noticeable peak shift is observed from 13.8° for the 2% Br film to 14.1° and 14.3° for the 45% and 85% Br films, respectively, providing clear evidence of Br incorporation into the crystal lattice substituting I-lattice site, in agreement with the PL findings. [32,33,37] Additionally, different peaks at 20°, 28°, and 32° in the 45% Br and 85% Br films suggest increased Br content in the perovskite-based composition [33,36,40].

To further assess and investigate the influence of the I/Br ratio on the morphology of perovskite films, top-view scanning electron microscopy (SEM) images were obtained for films with Br content of 2%, 45%, and 85% (Figure 1c-e). The SEM analysis reveals that all films maintain a compact, pinhole-free morphology, which is crucial for the fabrication of high-performance solar cells. Notably, as the Br content increases, a corresponding decrease in grain size is observed, implying that higher bromide content may enhance the rate of crystallization.[24] These morphological observations are corroborated by electrical simulations presented in the "Optical-electrical Modeling" section of the Supporting Information, which indicate a reduction in carrier lifetime at elevated bromide concentrations, an effect anticipated to limit the maximum attainable indoor PCE. Additionally, atomic force microscopy (AFM) measurements for the perovskite films (figure 1f-h) show surface roughness ($R_a$) values of 7.3, 8.4, and 10.2 nm for the samples with 2%, 45%, and 85% Br content, respectively. A rougher absorber layer is beneficial for light trapping and minimizing reflectance losses; however, it may also adversely affect device performance due to the presence of voids.[25,38-39,]

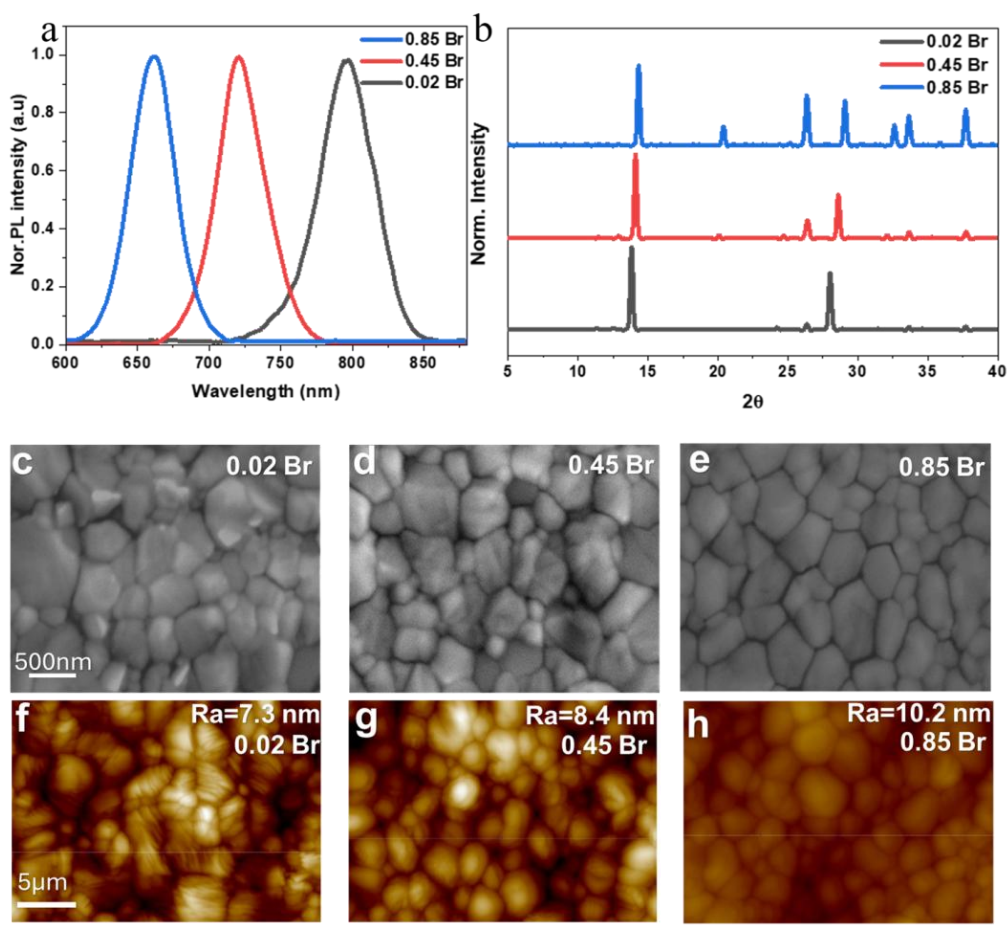

**Figure 1.** (a) PL, (b) XRD patterns, (c-e) SEM, and (f-h) AFM of perovskite films.

The thickness of the perovskite active layer, primarily governed by the concentration of the precursor solution[42-43], was investigated to determine the optimal parameters for PIPV devices. Specifically, devices utilizing the composition $FA_{0.90}Cs_{0.10}Pb(I_{0.98}Br_{0.02})_3$ ($E_g$ of 1.55 eV) were fabricated with precursor concentrations varying from 0.9 to 1.4 M. As detailed in Table S1, the PCE values were recorded as follows: 25.85% for 0.9 M, 27.07% for 1.1 M, 26.98% for 1.2 M, 30.05% for 1.3 M, and 29.46% for 1.4 M, measured at an illumination level of 1000 lux. The highest PCE of 30.05% was achieved at a precursor concentration of 1.3 M, which corresponded to a short-circuit current density ($Jsc$) of 121.4 $\mu$A/cm$^2$, an open circuit voltage ($Voc$) of 0.897 V, and a fill factor (FF) of 76.73% under illumination at 3000 K. The optimal layer thickness was determined to be within the range of 300-400 nm[27-29], as depicted in Figures 2a-c, which is consistent with the simulation results illustrated in Figure S3.

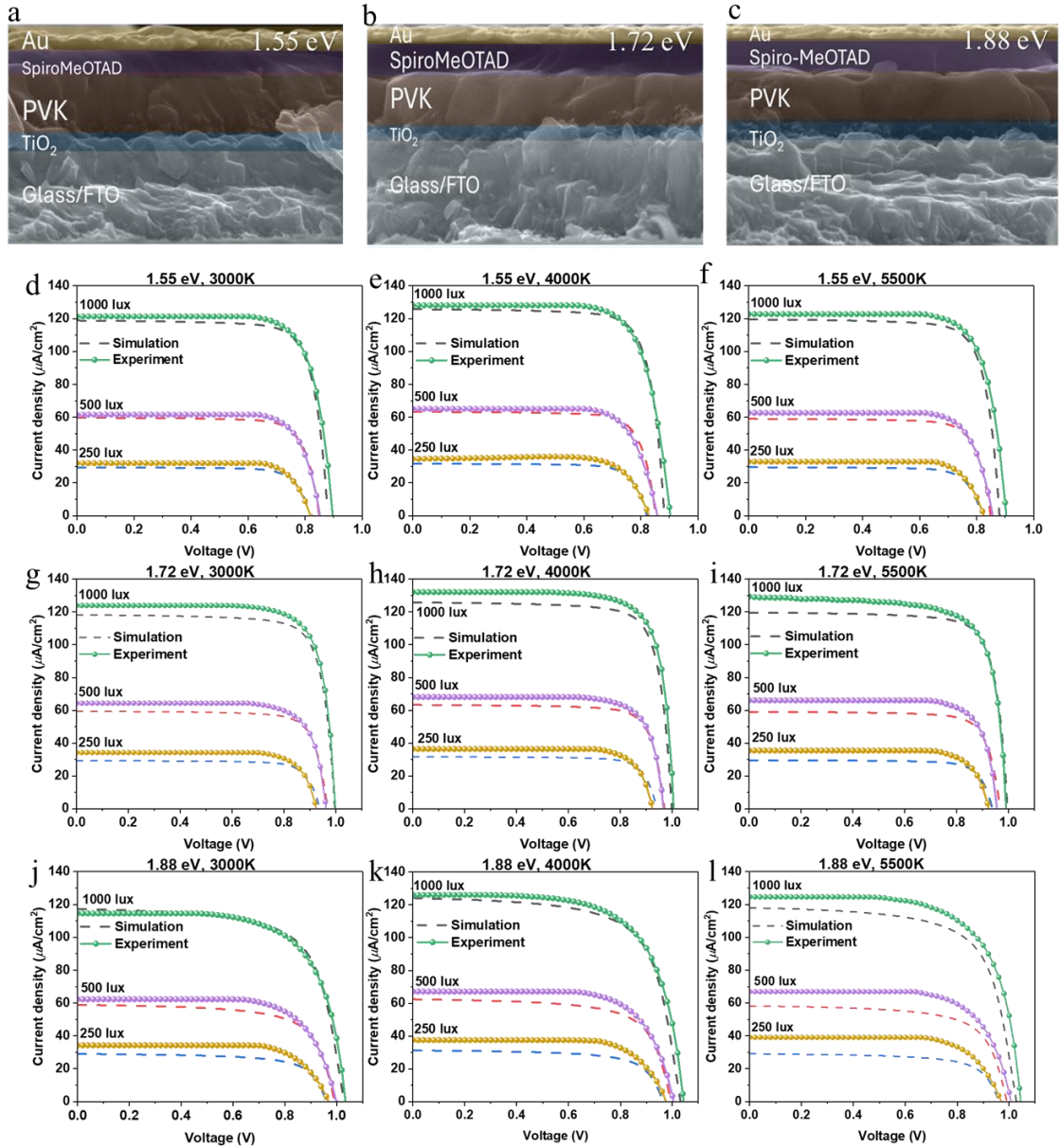

**Figure 2.** (a–c) Cross-sectional SEM images of corresponding cells with different bandgaps. (d–l) Experimental (solid) and simulated (dashed) *J–V* curves of perovskite solar cells with bandgaps of 1.55, 1.72, and 1.88 eV under 3000 K, 4000 K, and 5500 K indoor light (250–1000 lux).

After determining the optimal perovskite precursor concentration corresponding to film thickness, a constant concentration of 1.3 M was utilized for the 1.55 eV $E_g$, along with other compositions featuring $E_g$ of 1.72 eV and 1.88 eV. Subsequently, the solar cells were evaluated under various CTs of 3000 K, 4000 K, and 5500 K, as well as under light intensities of 1000, 500, and 250 lux, as depicted in Figure S2.

The devices with a 1.55 eV bandgap (Figure 2d-f, Table S2) demonstrated *Voc* ranging from 0.82 to 0.90 V, with the *Jsc* exhibiting a linear relationship with light intensity. At an illumination level of 1000 lux, *Jsc* was 128.1 $\mu$A/cm$^2$, while it decreased to 31.9 $\mu$A/cm$^2$ under dim lighting at 250 lux, indicating efficient current scaling. Among the varying color temperatures, the highest *Jsc* was recorded under 4000 K illumination, corresponding to a greater irradiance of 3.05 W/m$^2$. In comparison, reduced values were observed at both 3000 K and 5500 K due to diminished spectral overlap. The FF remained relatively constant, ranging from 75.0% to 79.7%, resulting in PCEs of 30% at 1000 lux (3000 K) and 31.3% at 250 lux (3000 K with an irradiance of 68 W/m$^2$). The PCEs were approximately 28% under both 4000 K and 5500 K conditions. Although the intermediate data at 500 lux followed similar trends (Table S2), the lower *Voc* limited efficiencies compared to higher bandgap devices. This outcome corroborates previous findings suggesting that lower band gaps are suboptimal for indoor conditions.[52,30] For the devices with a 1.72 eV bandgap (Figure 2g-i, Table S3), a higher current and voltage was observed. *Voc* values varied from 0.92 to 1.01 V, with *Jsc* reaching 132.1 $\mu$A/cm$^2$ under 1000 lux (4000 K), reducing to 33.5 $\mu$A/cm$^2$ under 250 lux. The FF remained almost constant, ranging from 74.6% to 78.5%. At 1000 lux, the efficiencies ranged from 32.2% to 35.0%, with a peak PCE of 35.04% achieved at 3000 K. At the level of 250 lux, the devices reached even higher efficiencies, achieving a PCE of 36.6% at 5500 K. The efficiency data at 500 lux mirrored these trends, maintaining values between 33.5% and 34.6% across all CTs. This bandgap (1.72 ev) exhibited reliable high efficiencies under various CTs and intensities, aligning with predictions that band gaps within the range of 1.70-1.80 eV are optimal for indoor photovoltaic applications[53-56] and consistent with reports indicating efficiencies exceeding 40% under LED lighting following passivation and compositional adjustments.[57-60]

The 1.88 eV devices (Figure 2j-l, Table S4) displayed the highest *Voc* values, ranging from 0.96 to 1.04 V. However, the current densities remained relatively low, peaking below 126 $\mu$A/cm$^2$, attributed to the reduced absorption of red photons. This phenomenon was most pronounced under 3000 K light, which contains a higher proportion of red wavelengths; however, it was less evident under 4000 K to 5500 K light, which is richer in blue photons.[61] The FF values for this bandgap were recorded between 67.4% and 74.9%. The PCE varied from 29.20% (3000 K, 1000 lux) to 37.44% (5500 K, 250 lux), marking one of the highest ever reported indoor efficiencies in the literature of PIPVs for devices utilizing an active area of $\geq$1 cm$^2$. Under lower illumination conditions of 250 lux (irradiance between 66 and 76 W/m$^2$), they presented superior performance under cooler white lighting, supporting findings from studies focused on optimizing perovskites for high CT light sources.[62]

To further explore these observations and understand practical constraints on PIPV performance, we conducted quantitative device simulations using optical full-wave electromagnetic modeling and electrical drift-diffusion modeling based on the Finite Element Method (see "Optical-Electrical Modeling" section in the SI).[72] Optical simulations provided spatially resolved charge-carrier generation profiles for various WLEDs CTs (3000, 4000, and 5500 K), illumination intensities (1000, 500, and 250 lux), and perovskite bandgaps (1.55, 1.72, and 1.88 eV). These generation profiles were input into the electrical model to solve the steady-state Poisson and continuity equations, yielding the simulated *J–V* curves in Figure 2d-l.

To ensure quantitative agreement with experimental data, the model incorporated Shockley–Read–Hall trap-assisted recombination,[73–75] where carrier lifetimes were calibrated against experimental data for each bandgap,[76–78] yielding values ranging from 20 to 55 ns. These values align with several theoretical and experimental studies on IPVs.[74,76,79] Moreover, the simulated $J$–$V$ curves (dashed) show good agreement with experimental data (solid), with modeled $J_{SC}$, $V_{OC}$, FF, and maximum power output following the experimental trends as a function of CT and illumination level.

Slight underestimation of $J_{SC}$ in simulations is primarily due to the abrupt absorption cutoff at the bandgap wavelength (used to avoid inconsistencies in sub-bandgap optical data across materials). Additionally, the rougher surfaces observed in the experimental devices (Figure 2a-c) also enhance absorption compared to the idealized planar interfaces assumed in the model.[80] Importantly, the extracted carrier lifetime for the 1.88 eV devices was approximately 20 ns, compared to ~50 ns for the 1.55 and 1.72 eV cases, leading to a reduction in maximum attainable $V_{OC}$, which produces higher $Voc$ deficits. This shorter lifetime correlates with the smaller grain sizes observed at higher bromide content (see Figure 1), indicating reduced carrier lifetime as a key limiting factor in the maximum attainable performance of wide-bandgap PIPVs. One way to overcome such problems and further improve indoor PCEs in the future might be to try optimizing and increase the grain size of wide band gap perovskites. This could lead to indoor PCEs close to the efficiency limit.

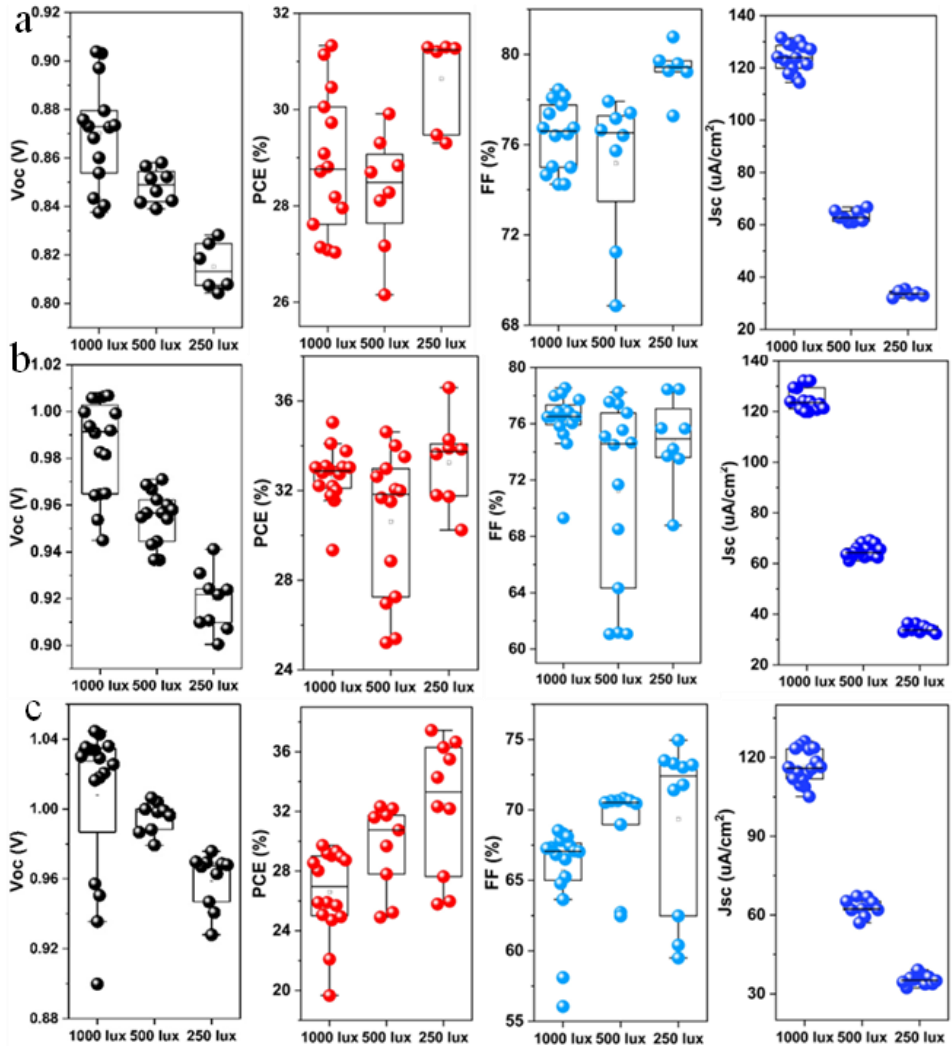

**Figure 3.** Statistical analysis of the device's performance with (a) 1.55, (b) 1.72, and (c) 1.88 eV.

Statistical analysis of the device performance reveals significant differences among the three examined band gaps. The devices with a bandgap of 1.55 eV (Figure 3a) exhibited photocurrents of $123.9 \pm 5.4$ $\mu$A/cm$^2$ at 1000 lux, alongside fill factors of $76.6 \pm 1.5\%$ and a *Voc* of $0.87 \pm 0.02$ V. The corresponding efficiencies recorded were $28.7 \pm 1.5\%$ at 1000 lux, $28.5 \pm 1.2\%$ at 500 lux, and $31.2 \pm 1.0$ % at lower illumination levels (250 lux). In contrast, the devices with a 1.72 eV bandgap (Figure 3b) demonstrated a superior performance profile, achieving a higher *Voc* of $0.99 \pm 0.02$, a *Jsc* of $123.5 \pm 4.5$ $\mu$A/cm$^2$ at 1000 lux, and a consistent FF of $76.0 \pm 2.0$ %. This configuration resulted in efficiencies of 32.8 ± 1.2% at 1000 lux, 31.8 ± 3.2% at 500 lux, and 33.7 ± 1.9% at 250 lux. Conversely, the 1.88 eV absorber attained the highest *Voc*, with efficiencies recorded at 27 ± 2.9% (1000 lux), 30.8 ± 2.9% (500 lux), and peaking at 33.3 ± 4.5% (250 lux). Overall, the findings substantiate the conclusion that the 1.72 eV composition provides a more reproducible performance across varying light conditions, while the 1.88 eV band gap variant provided the highest indoor efficiency of ~ 37% at low light intensity conditions.

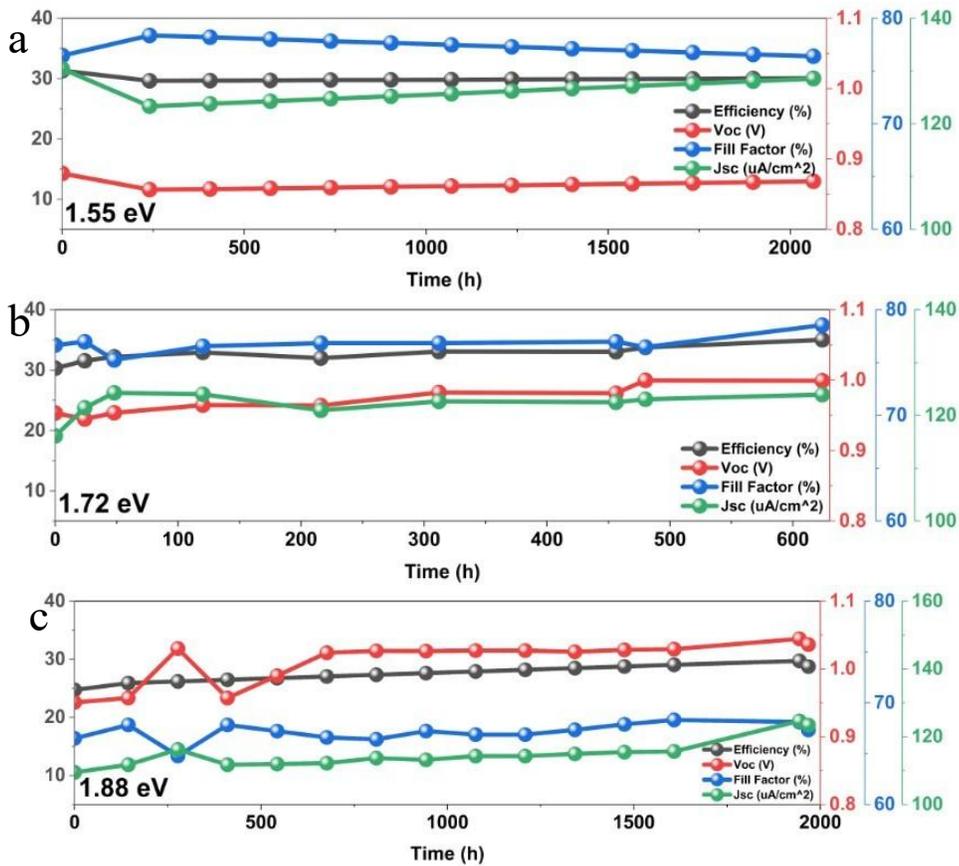

**Figure 4.** The stability of the devices with (a) 1.55, (b) 1.72, and (c) 1.88 eV.

Finally, stability assessments were conducted at an ambient temperature of 25 ± 5°C, under relative humidity conditions of 30 ± 5%, and with illumination at 1000 lux, as depicted in Figure 4a. The performance of the $FA_{0.85}Cs_{0.15}Pb(I_{0.55}Br_{0.45})_3$ (1.72 eV) device demonstrated notable improvements over a period exceeding 600 hours. Specifically, the PCE increased from 30.31% to 35.04%, accompanied by an enhancement in *Voc* from 0.954 V to 0.999 V. The FF remained stable within the range of 77% to 78.5%, while the *Jsc* stabilized at approximately 124 $\mu$A/cm$^2$. In parallel, the FA$_{0.85}$Cs$_{0.15}$Pb(I$_{0.45}$Br$_{0.55}$)$_3$ (1.88 eV) composition, illustrated in Figure 4c, exhibited a similar performance trajectory over 2000 hours, with its PCE increasing from 25% to 29%, *Voc* rising from 0.94 V to 1.04 V, *Jsc* improving from 114 to 124 $\mu$A/cm$^2$. FF increasing from 58% to 68%. Conversely, the FA$_{0.90}$Cs$_{0.10}$Pb(I$_{0.98}$Br$_{0.02}$)$_3$ (1.55 eV) device, as shown in Figure 4a, maintained stable performance throughout the 2000-hour duration, sustaining a PCE in the range of 29% to 30%. The observed stability of the 1.72 eV and 1.88 eV compositions presents an encouraging prospect for their application in indoor lighting.

**Conclusion**

This study investigated the compositional engineering of formamidinium-cesium-based perovskite Cs$_x$FA$_{x-1}$Pb(I$_{1-y}$Br$_y$)$_3$ for IPV applications. Through photoluminescence (PL) spectroscopy, three targeted bandgap energies of 1.55, 1.72, and 1.88 eV were successfully realized. Meanwhile, X-ray diffraction (XRD) revealed the effective incorporation of bromine into the crystal lattice, which surprisingly presented a suppressed $\delta$-phase and PbI$_2$ impurities, which are reported systematically for indoor solar cell applications. Morphological analysis using scanning electron microscopy (SEM) and atomic force microscopy (AFM) revealed the formation of dense, pinhole-free films. However, a higher bromine content led to a reduced grain size and increased surface roughness due to rapid crystallization. On one hand this helped the devices to achieve high *Jsc* but on the other hand it was identified (from simulations) as a limiting factor towards attaining the maximum possible indoor PCE. An optimal active layer thickness of 300 to 400 nm was established by using a precursor concentration of 1.3 M. Performance tests under different light intensities from white light-emitting diodes (WLEDs) revealed that among the wide band gap devices the 1.72 eV bandgap composition had the highest $Jsc$, while the 1.88 eV band gap presented the highest $Voc$. The device with a 1.88 eV bandgap achieved a notable indoor PCE of 37.4% at low light intensity (250 lux), surpassing the performance of other compositions. The 1.72 eV device, characterized by high values of $Jsc$, open circuit voltage ($Voc$), and fill factor (FF), achieved PCEs of 36.6% under 250 lux and 35.04% under 1000 lux, establishing it as the optimal candidate for indoor energy harvesting applications over different indoor illumination conditions (intensity and color temperature).

**ACKNOWLEDGMENT**

E. A. A, M.S.A and gratefully acknowledges the support from King Abdulaziz City for Science and Technology (KACST), Saudi Arabia. A.H.S and I.B.K acknowledge the support from a research grant funded by the Research, Development, and Innovation Authority (RDIA) – Kingdom of Saudi Arabia – with grant number (12979-iau-2023-TAU-R-3-1-EI-). G.K, N.T, K.P and E.K gratefully acknowledge the support from the action: "Promotion of quality, innovation and extroversion in universities (ID 16289)", "SUB1.1 Clusters of Research Excellence - CREs" and funded by the Special Account of the Ministry of Education, Religious Affairs and Sports within the framework of the National Recovery and Resilience Plan "Greece 2.0", with funding from the European Union – NextGenerationEU and co-financing from national resources (National Public Investments Program – VAT contribution). G.P and M.K gratefully acknowledge the Hellenic Foundation for Research and Innovation (HFRI) under "Sub-action 2 for Funding Projects in Leading-Edge Sectors - RRFQ: Basic Research Financing (Horizontal support for all Sciences)", MultiCool (15117). F.I.A acknowledges the support from King Saud University, Saudi Arabia.

## Author Contributions

E.A.A conceived the idea of the work, designed, planned the experiments and supervised the work. M.S.A, F.I,A, T.F.A, and A.S.A fabricated and optimized the perovskite solar cell devices, did all the basic characterizations, analyzed the data and wrote the first draft with support from E.A.A, G.K, T.F.A and N.R.A. E.A.A, G.K and G.P revised the first draft and contributed to the explanation of the results. G.P performed the simulations and wrote the simulations part. H.A, A.H.S and I.H.K contributed to the results, discussion and work supervision. N.T, D.T, M.M.A and A.A were responsible for top-view SEM, cross-section AFM, PL measurement and analysis. All authors contributed towards the preparation of the manuscript and approved its submission.

## Experimental and Methods

### Materials

The perovskite solar cells were prepared using one-step method and the structure is mesoscopic n-i-p architecture. Titanium diisopropoxide bis(acetylacetonate) solution (75% in 2-propanol) and Dimethylformamide (DMF) and Dimethyl Sulfoxide (DMSO) and Acetonitrile and Ethanol and isopropanol and Lithium bis(trifluoromethane)sulfonimide was purchased from Sigma Aldrich. mesoporous $TiO_2$ paste (30NRD) and Methylammonium chloride (MACl) and Formamidinium Iodide (FAI) were purchased from Greatcell Solar Materials Pty Ltd. Cesium iodide(CsI) and lead iodide ($PbI_2$ and lead bromide ($PbBr_2$) were purchased from alfa acer. 2,2',7,7'-Tetrakis[N,N-di(4methoxyphenyl)amino]-9,9'-spirobifluorene(spiroOMe-TAD) was purchased Xi'an Polymer Light Technology Corp. Gold (Au) was purchased from Plasmaterials.

**Device Fabrication**

Fluorine-doped tin oxide (FTO) glass substrates measuring 2.5 × 2.5 cm (TCO glass, NSG 10, Nippon Sheet Glass, Japan) underwent a series of surface treatments. Initially, the substrates were etched using a mixture of zinc powder and 4 M hydrochloric acid. Subsequent cleaning involved ultrasonication in a 2% Hellmanex solution followed by thorough rinsing with deionized water and ethanol. To further prepare the surfaces, the substrates were subjected to oxygen plasma treatment for 15 minutes. Titanium dioxide ($TiO_2$) was then deposited onto the cleaned FTO substrates via spray pyrolysis at a temperature of 450 °C. The precursor utilized was a commercial titanium diisopropoxide bis(acetylacetonate) solution, which was 75% concentrated in 2-propanol and diluted with anhydrous ethanol in a 1:9 volume ratio; oxygen served as the carrier gas. Additionally, a mesoporous $TiO2$ layer was fabricated through spin-coating a diluted paste in ethanol, following a 1:6 weight ratio, at a rotation speed of 5000 rpm for 15 seconds, and subsequently sintered at 450 °C for 30 minutes in a dry atmosphere. Perovskite films were deposited utilizing a single-step deposition method. The precursor solution was prepared within a nitrogen atmosphere, incorporating MACl, CsI, FAI, $PbI_2$, and $PbBr_2$ dissolved in anhydrous dimethylformamide/dimethyl sulfoxide at a 4:1 volume ratio to achieve the desired compositions: $FA_{0.85}Cs_{0.15}Pb(I_{0.15}\text{-}Br_{0.85})_3$, $FA_{0.85}Cs_{0.15}Pb(I_{0.55}\text{-}Br_{0.45})_3$ $FA_{0.90}Cs_{0.10}Pb(I_{0.98}\text{-}Br_{0.02})_3$. These perovskite films were subsequently passivated by dynamically spin-coating a solution containing 3 mg of n-Octyl ammonium Iodide (OAI) in 1 mL of isopropanol at a speed of 4000 rpm for 30 seconds. The hole transport material (HTM) was applied by spin-coating at 4000 rpm for 25 seconds. The spiro-OMe-TAD was doped with bis(trifluoromethylsulfonyl)imide lithium salt using 24 µL of a solution comprised of 520 mg of LiTFSI in 1 mL of acetonitrile, along with 28.8 μL of 4-tert-butylpyridine. Finally, an approximately 80-100 nm layer of gold (Au) was deposited by thermal evaporation, completing the device fabrication.

**Scanning Electron Microscopy (SEM):**

The morphological characteristics of the perovskite samples were investigated using a field-emission scanning electron microscope (JSM-6010PLUS/LV, JEOL Ltd.) at an operating voltage of 10 keV.

**Atomic force microscopy (AFM):**

The surface morphology was studied using the Bruker dimension ICON AFM system.

**X-ray diffraction (XRD):**

The crystal structure was evaluated using X-ray diffraction with a Rigaku Mini Flex apparatus, with a scanning range set from 10° to 50°.

**Steady-state photoluminescence (PL):**

Spectra were acquired using a Horiba FL3C-222 system, with excitation at a wavelength of 450 nm.

**The current-voltage (J-V) characteristics:**

J–V measurements of indoor PVK cells were recorded under indoor light using the ILS-30 standard spectrum simulator (Enlitech Technology). LEDs were stabilized for 30 min before recording the measurements. The devices were measured at 3000, 4000, and 5000 K under 1000 lux, 500 lux, and 250 lux. The aperture mask area was 1 cm².

## Optical-Electrical Modeling

The $J$–$V$ characteristics and photovoltaic (PV) parameters ($V_{OC}$, $J_{SC}$, FF, and PCE) of the examined perovskite-based solar cells (PSCs) were simulated using optical full-wave electromagnetic simulations and electrical drift–diffusion modelling based on the Finite Element Method.[72]

Optical simulations were performed by numerically solving Maxwell's equations in the frequency domain to compute the total electric field vector $\boldsymbol{E}(z, v)$ at each position $z$ in the PSC and at each frequency $v$. From $\boldsymbol{E}(z, v)$, the spectral absorbed energy density $Q(z, v)$ in the photoactive layer is calculated as:[81]

$$Q(z, v) = \frac{1}{2} c \varepsilon_0 \alpha(v) n(v) \boldsymbol{E}(z, v)^2, (1)$$

where $c$ is the vacuum speed of light, $\varepsilon_0$ is the vacuum permittivity, $\alpha = 2\pi\kappa/\lambda$ is the absorption coefficient, with $n$ and $\kappa$ representing the refractive index and extinction coefficient (i.e., real and imaginary part of the complex refractive index) of the photoactive layer, respectively. The absorbed energy density $Q(z, v)$ is then used to determine the spatially resolved spectral electron-hole generation rate $G(z, v)$:[81,82]

$$G(z, v) = \frac{Q(z, v)}{hv} = \frac{\pi \varepsilon'' \varepsilon_0}{h} \boldsymbol{E}(z, v)^2, (2)$$

where $h$ is Planck's constant and $\varepsilon''$ (=$2nk$) is the imaginary part of the permittivity of the photoactive layer. The total spatially resolved generation rate $G(z)$ is obtained by integrating $G(z, v)$ over frequency $v$, weighted by the WLED illumination spectrum $I_{WLED}(v)$, corresponding to various color temperatures (CTs) and intensities (see Figure S2):[73–76,81,83]

$$G(z) = \int G(z, \nu) I_{WLED}(\nu) d\nu \, , (3)$$

Based on $G(z)$, electrical simulations were performed by numerically solving the steady-state Poisson and continuity equations to obtain the electron and hole current densities. For the one-dimensional (1D) case, these equations are explicitly given by:[74–76,83]

$$\frac{d^2\psi(z)}{dz^2} = \frac{q}{\varepsilon_0 \varepsilon_r}(n(z) - p(z) + N_A - N_D), (4)$$

$$\frac{1}{q}\frac{dJ_n}{dz} + G(z) - U(z) = 0, (5)$$

$$\frac{1}{q}\frac{dJ_p}{dz} + U(z) - G(z) = 0, (6)$$

where $\psi$ is the electrostatic potential, $q$ is the elementary charge, $\varepsilon_0$ and $\varepsilon_r$ are the vacuum and relative permittivity, $n$ and $p$ are electron and hole concentrations, $N_A$ and $N_D$ are the acceptor and donor doping concentrations, and $J_n$ and $J_p$ are the electron and hole current densities, respectively, described by the drift–diffusion constitutive relations.[73,74,76,83] In eqs 5 and 6, $G(z)$ is the generation rate (see eq 3), and $U(z)$ is the recombination rate, modeled using Shockley–Read–Hall (SRH) nonradiative trap-assisted recombination assuming equal carrier lifetimes for electrons and holes.[73,74,83]

The device architecture used in this study is FTO/TiO$_2$/Cs$_x$FA$_{x-1}$Pb(I$_{1-y}$Br$_y$)$_3$/Spiro-OMeTAD/Au (see also Figure 2a–c), where fluorine-doped tin oxide (FTO) and gold (Au) serve as the transparent and metal electrodes, respectively. Titanium dioxide (TiO$_2$) and Spiro-OMeTAD form the electron-transporting layer (ETL) and hole-transporting layer (HTL), respectively. Cs$_x$FA$_{x-1}$Pb(I$_{1-y}$Br$_y$)$_3$ is the photoactive perovskite layer. The complex refractive indices of perovskites films with different bandgaps were obtained from refs 84–86. Material parameters used in the electrical simulations are listed in Table S1 and correspond to typical indoor PSC operation conditions.[75–78,83] Electron-hole carrier lifetimes for each bandgap were calibrated against experimental data to ensure accurate simulation results.[75,77,78] The optical-electrical model was calibrated under 3000 K and 1000 lux WLED illumination for the three specific bandgaps: 1.55, 1.72, and 1.88 eV. Finally, the calibrated PSCs were evaluated under different WLED CTs (3000, 4000, and 5500 K) and varying illumination intensities (1000, 500, and 250 lux).

**Table S1.** Material parameters used in the device simulations, where CB and CB are the conduction and valence band, respectively.[75–78,83]

| Parameters | TiO$_2$ | Cs$_x$FA$_{x-1}$Pb(I$_{1-y}$Br$_y$)$_3$ | Spiro-OMeTAD |
|---|---|---|---|
| Thickness (nm) | 200 | 100–1000 | 200 |
| Bandgap (eV) | 3 | 1.55–1.85 | 3 |
| Electron affinity (eV) | 4 | 3.9 | 2.35 |
| Dielectric permittivity | 10 | 6.5 | 3 |
| CB density of states (cm$^{-3}$) | $1\times10^{20}$ | $1.8\times10^{18}$ | $2.2\times10^{18}$ |
| VB density of states (cm$^{-3}$) | $1\times10^{20}$ | $1.8\times10^{18}$ | $1.8\times10^{19}$ |
| Electron mobility (cm$^2$/V·s) | 20 | 2 | $2\times10^{-4}$ |
| Hole mobility (cm$^2$/V·s) | 10 | 2 | $2\times10^{-4}$ |
| Shallow uniform donor density (cm$^{-3}$) | $1\times10^{18}$ | 0 | 0 |
| Shallow uniform acceptor density (cm$^{-3}$) | 0 | $1\times10^{14}$ | $2\times10^{18}$ |
| Electron lifetime (ns) | 5 | 20–55 | 5 |
| Hole lifetime (ns) | 5 | 20–55 | 5 |

**Supporting Information**

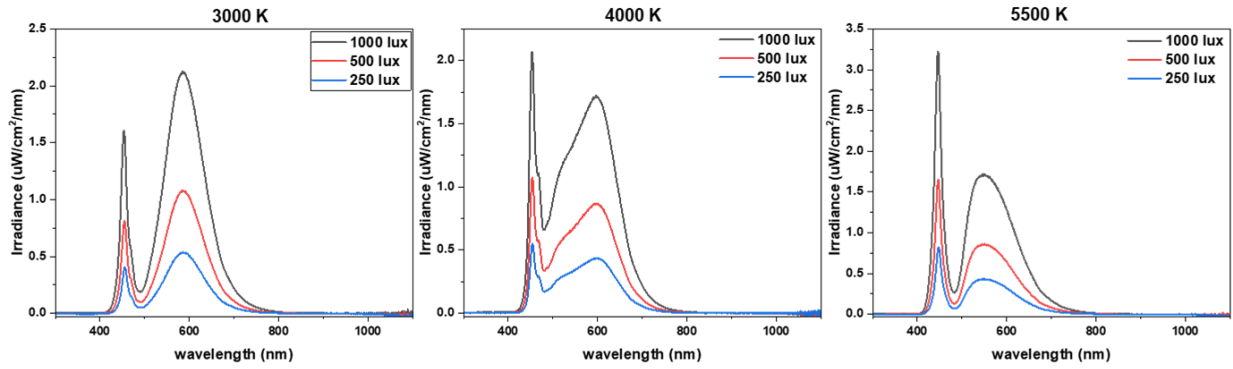

**Figure S1.** Spectrum of WLED with various color temperatures (see titles) and intensities (see legends).

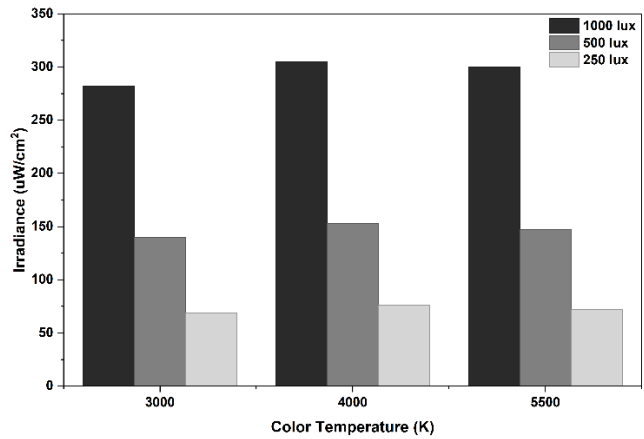

**Figure S2.** Irradiance levels under different color temperatures (3000 K, 4000 K, 5500 K) at various illuminance conditions (1000, 500, and 250 lux).

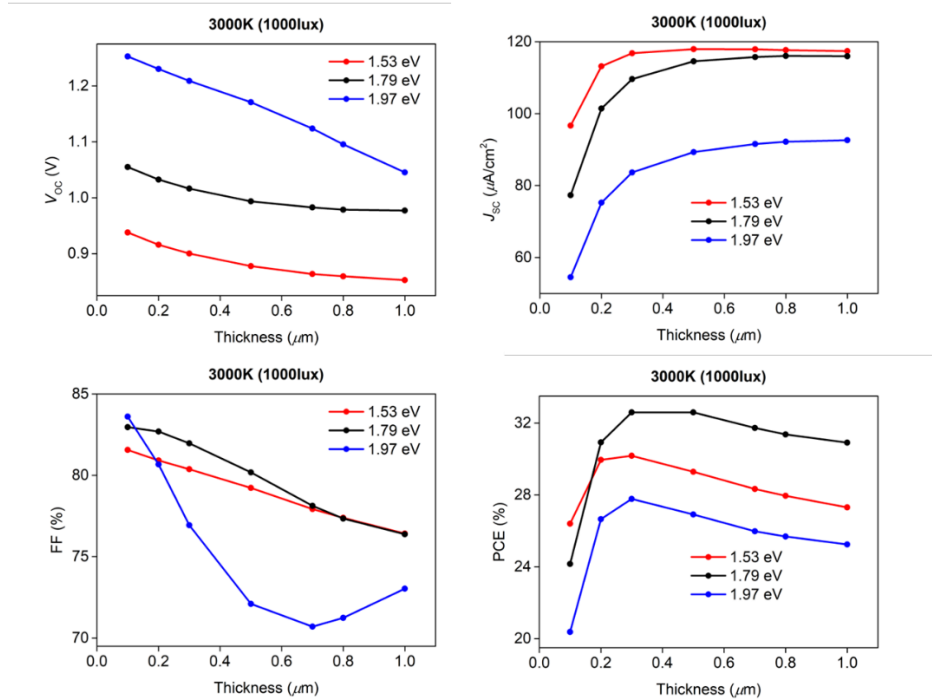

**Figure S3.** Simulation of photovoltaic parameters (*Voc*, *Jsc*, FF, and PCE) for perovskite solar cells with varying bandgaps as a function of perovskite layer thickness under 3000 K and 1000 lux illumination.

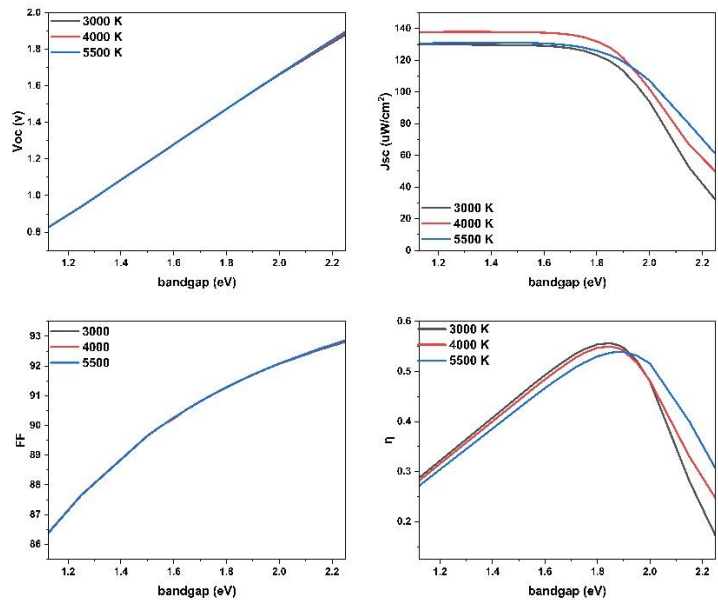

**Figure S4.** Theoretical SQ limit analysis of solar cell performance under WLED with various color temperatures.

| Precursor concentration (M) | Voc (V) | PCE (%) | Fill Factor (%) | Jsc ($\mu$A/cm$^2$) |
|---|---|---|---|---|
| 0.9 | 0.831 | 25.85 | 76.15 | 114.8 |
| 1.1 | 0.837 | 27.07 | 73.83 | 123.1 |
| 1.2 | 0.854 | 26.98 | 72.05 | 122.8 |
| 1.3 | 0.897 | 30.05 | 76.73 | 121.4 |
| 1.4 | 0.802 | 29.46 | 78.99 | 130.2 |

**Table S1.** Performance of the 1.55 eV devices for different precursor concentrations.

| K/Lux | Voc (V) | PCE (%) | Fill Factor (%) | Jsc ($\mu$A/cm$^2$) |
|---|---|---|---|---|
| 3000-1000 | 0.897 | 30.05 | 76.73 | 121.365 |
| 3000-500 | 0.852 | 29.91 | 77.92 | 61.568 |
| 3000-250 | 0.819 | 31.29 | 79.72 | 32.000 |
| 4000-1000 | 0.903 | 28.81 | 75.03 | 128.127 |
| 4000-500 | 0.858 | 28.28 | 75.73 | 65.081 |
| 4000-250 | 0.828 | 29.30 | 77.27 | 34.601 |
| 5500-1000 | 0.904 | 28.72 | 76.74 | 122.765 |
| 5500-500 | 0.857 | 28.11 | 77.16 | 62.619 |
| 5500-250 | 0.825 | 29.47 | 79.23 | 32.944 |

**Table S2.** Performance table of the 1.55 eV devices across different color temperatures and intensities.

| K/Lux | Voc (V) | PCE (%) | Fill Factor (%) | Jsc ($\mu$A/cm$^2$) |
|---|---|---|---|---|
| 3000-1000 | 0.999 | 35.04 | 78.54 | 123.943 |
| 3000-500 | 0.962 | 34.61 | 77.42 | 64.373 |
| 3000-250 | 0.924 | 36.59 | 78.46 | 34.262 |
| 4000-1000 | 1.006 | 34.10 | 78.17 | 132.107 |
| 4000-500 | 0.967 | 33.51 | 77.56 | 68.134 |
| 4000-250 | 0.924 | 34.27 | 78.44 | 36.448 |
| 5500-1000 | 0.992 | 32.22 | 74.60 | 129.452 |
| 5500-500 | 0.957 | 34.01 | 78.24 | 66.094 |
| 5500-250 | 0.924 | 36.61 | 77.93 | 35.505 |

**Table S3.** Performance table of the 1.72 eV devices across different color temperatures and intensities.

| K/Lux | $Voc$ (V) | PCE (%) | Fill Factor (%) | $Jsc$ ($\mu$A/cm$^2$) |
|---|---|---|---|---|
| 3000-1000 | 1.034 | 29.20 | 68.54 | 114.56 |
| 3000-500 | 0.998 | 32.19 | 70.82 | 62.30 |
| 3000-250 | 0.963 | 36.65 | 74.95 | 33.66 |
| 4000-1000 | 1.043 | 29.31 | 67.45 | 126.09 |
| 4000-500 | 1.004 | 31.75 | 70.65 | 67.14 |
| 4000-250 | 0.976 | 36.29 | 73.02 | 37.55 |
| 5500-1000 | 1.044 | 29.72 | 68.11 | 124.63 |
| 5500-500 | 1.006 | 32.33 | 70.66 | 66.93 |
| 5500-250 | 0.969 | 37.44 | 71.78 | 39.09 |

**Table S4.** Performance table of the 1.88 eV devices across different color temperatures and intensities.